# Solar Energetic Particle Events and Radio Bursts

Nat Gopalswamy[1]

[1] NASA Goddard Space Flight Center, Maryland, US

nat.gopalswamy@nasa.gov

**Abstract.** Solar Energetic Particles (SEPs) and radio bursts are indicators of particle acceleration on the Sun and in the heliosphere. The accelerated particles have energies significantly higher than thermal particles up to several orders of magnitude. SEPs are detected directly by particle detectors on Earth and in space. Understanding SEPs is important from both science and application points of view because they are poorly understood and present space weather hazard to humans and their technology in space. SEPs accompany energetic flares, coronal mass ejections (CMEs), and intense radio bursts, which help us understand particle properties such as intensity, spectra, and time evolution. This paper summarizes how SEP properties are closely related to solar eruptions and the associated solar radio bursts.

## 1 Introduction

Solar energetic particle (SEP) events and radio bursts are indicative of particle energization in the corona and interplanetary (IP) medium. During solar eruptions, the Sun releases large amounts of stored energy in solar magnetic regions such as active regions and quiescent filament regions. Solar flares and coronal mass ejections (CMEs) are two of the key manifestations of solar eruptions. Eruptive prominences and fast mode magnetohydrodynamic (MHD) shocks form the inner core and outermost structure of CMEs, respectively. A significant amount of the released energy is converted into the kinetic energy of the energetic particles that manifest as SEPs and various types of radio bursts. SEPs were first observed in 1942 by ground-based instruments used for detecting cosmic rays [1]; they are now routinely observed by spaceborne detectors. The ground level enhancement (GLE) events are SEP events that indicate tens of GeV particles. In space, SEP ions are observed at energies down to the solar wind particle energies [2-3]. SEPs also consist of electrons, which are commonly observed at energies up to a few MeV [4], although they have been observed up to a few 100 MeV [5]. Indirect observations of energetic particles are via their electromagnetic signatures. Interestingly, nonthermal radio emission from the Sun were also first detected in 1942 [6]. While all radio emission from the Sun is due to electrons, energetic ions produce gamma-ray emission when the particles interact with the dense solar atmosphere. Particles are energized in the solar corona and propagate toward and away from the Sun. Electrons propagating away from the Sun generate various types of radio bursts [7] at wavelengths ranging from decimeters to kilometers. Electrons flowing toward the Sun produce microwave bursts from cm to mm wavelengths. Some of these electrons also produce hard X-ray bursts and continuum emission extending to 100s of MeV [8]. The energetic electrons responsible for the radio bursts and SEP events have a common source of energization. Therefore, radio bursts are often used to infer SEP events in advance. While SEP events significantly alter Earth's space environment resulting in adverse impact on humans and their technology in space, radio bursts generally do not have direct space weather effect. There is one exception: occasional intense microwave emission can interfere with spacecraft signals (GPS) and radar signals (airport operations) [9-10]. Extensive discussion on solar energetic particles and radio bursts can be found in the monographs [11-12]. Recent reviews on SEPs can be found in [13-14]. This article focuses on large SEP events relevant for space weather (those with a >10 MeV flux of at least 10 pfu (particle flux units), and the associated radio bursts.

## 2      Why do we care about SEPs?

SEPs represent one of the major radiation hazards in space, both inside Earth's magnetosphere and outside in the IP space. According to NOAA's Space Weather Prediction Center, large SEP events are of concern for (i) high frequency (HF) radio communication systems, (ii) satellite operations, and (iii) biological systems [15]. The SEP events are called "Solar Radiation Storms" whose severity is classified on a scale of 1-5 denoted by S1-S5 to indicate >10 MeV proton intensity increasing by a factor 10 from10 pfu to $10^5$ pfu. S1 events have a minor impact on HF radio in the polar regions with little impact on satellite operations and biological systems. The severity of the impact progressively increases for S2, S3, S4, and S5 levels on all aspects (i)-(iii). In the extreme case of ~$10^5$ pfu (S5), a complete HF radio blackout is possible; wide-ranging problems may arise in satellite operations – loss of mission, loss of control, permanent damage to solar panels, and image degradation due to noise created in the detectors; high radiation risk to passengers and crew in polar aircraft and astronauts performing extravehicular activities.

There is plenty of anecdotal evidence showing extensive damage to space missions or subsystems due to particle radiation. The widespread impact of the 2003 Halloween storms has been well documented [16]. The Martian Radiation Environment Experiment (MARIE) on the Mars Odyssey mission was designed to assess the radiation environment of Mars. MARIE succumbed to the SEP event that occurred on 28 October 2003 [17]. Damage to solar panels of satellites has also been well documented, the main concern being the reduction of efficiency. Two powerful SEP events that occurred on 29 September 1989 and October 19 each produced a step-like decrease in the expected current from the GOES-7 solar panels by ~5–10% [18]. Apart from these super-intense SEP events, regular SEP events result in frequent spacecraft anomalies. Statistical studies have shown that the frequency of spacecraft anomalies depends on the intensity and energy range of SEP events, the spacecraft altitude, and spacecraft inclination [19]. A notable result is that the number of spacecraft anomalies increases after SEP events and peaks 4–5 days following the start of an SEP event. Even though there is not much advance warning for GLE events, one can take precautions when an SEP event occurs owing to the possibility of spacecraft anomalies over the following few days. These results illustrate the importance of predicting the intensity, temporal evolution, and spectrum of SEP events [20].

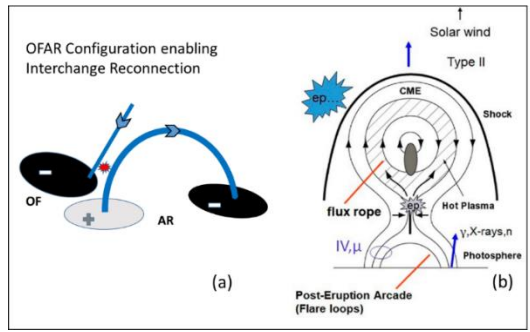

**Fig. 1.** Particle acceleration sites in the corona: (a) interchange reconnection region formed between an open field (OF) region and a closed magnetic field region such as an active region (AR). Such a magnetic configuration is known as an OFAR region. The interchange reconnection region is denoted by the red patch. (b) Flare reconnection in a closed magnetic region that produces post eruption arcade and a coronal mass ejection (CME). Particles are energized in the reconnection region and at the shock (ep denotes acceleration of electrons and protons). Electrons accelerated in the flare reconnection region produce microwave (µ) type IV bursts, and hard X-ray bursts ("X-rays"). Accelerated protons precipitating from the flare site result impulsive gamma-ray bursts ("γ") and neutron ("n") emission while interaction with the solar atmospheric particles. The shock-accelerated electrons and ions are observed as SEPs in space; the lower energy electrons produce type II bursts via the plasma emission mechanism.

## 3   Sources of energetic particles

There are many sources of energetic particles in the heliosphere [21]. Here we discuss those sources involving energy release at the Sun. There are other sources of energetic particles that we do not discuss here: reconnection exhausts in the solar wind [22], stream interaction regions [23-24], and planetary atmospheres [25-26].

The commonly accepted sources of energetic particles from the Sun are shown schematically in Fig. 1, viz., interchange reconnection between open and closed field lines (Fig.1a) and reconnection between closed field lines (Fig. 1b). In both cases a closed magnetic region such as an active region (AR) or a quiescent filament region is involved. The magnetic region resulting in interchange reconnection has an open filed + AR (OFAR) magnetic configuration. OFAR configuration is thought to be responsible for type III storms, which consist of short duration type III-like bursts that occur in rapid succession for days, often lasting for more than one solar rotation [27-30]. The bursts in type III storms are caused by energetic electrons accelerated in the interchange reconnection region. A similar configuration is involved in causing impulsive SEP events, but the configuration is more dynamic, involving jets [31]. These events may also appear as narrow CMEs in the coronagraph field of view [32]. The helium isotope

$^3$He is significantly enhanced in these events, and hence are known as $^3$He-rich events. One of the primary characteristics of impulsive SEP events is the association with type III radio bursts. These are regular type III bursts that are more intense and last longer than the type III storm bursts. The type III burst association indicates that electrons are also accelerated in the source region of impulsive SEP events.

Figure 1b schematically shows particle energization sites during solar eruptions. The energy release is thought to be due to the so-called flare reconnection in the current sheet below the rising flux rope. The flare reconnection results in two magnetic structures, viz., the post eruption arcade (PEA) at the Sun and the ejected flux rope. The flux rope accelerates while the reconnection is in progress and can attain high speeds, occasionally exceeding 3000 km/s. When the flux rope speed exceeds the local magnetosonic speed, a fast mode MHD shock is formed ahead of the flux rope. The shock can accelerate protons and electrons to very high energies [33-34]. Large SEP events are thought to be primarily due to shock acceleration [2], although there is an ongoing debate on the contribution from the associated flare [35-36]. The two sites of particle acceleration in Fig. 1b are evidenced by confined flares (or compact flares [37]) that do not have an associated CME and CME-driven shocks from quiescent filament region that have only weak flare signatures but are associated with large SEP events.

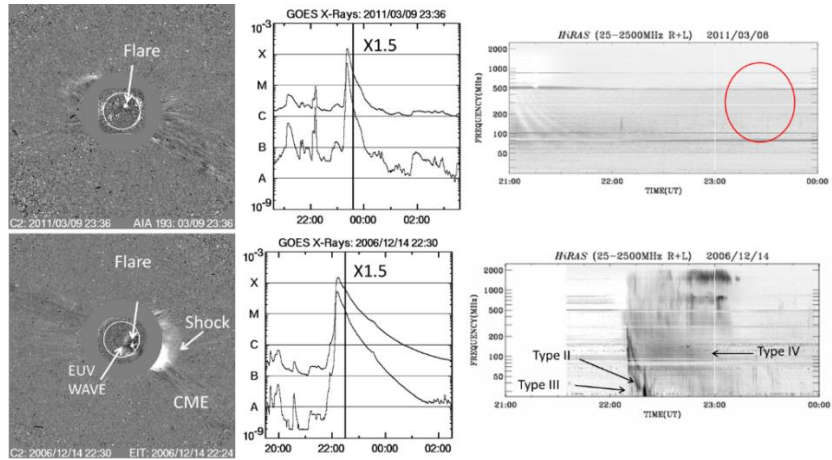

**Fig. 2.** (top left) Coronagraph image showing a flare brightening in an EUV difference image (SDO/AIA 193 Å) superposed on a LASCO/C2 difference image. (top middle) GOES soft X-ray light curve in the 1-8 Å band showing the X1.5 flare. The vertical line marks the time of the coronagraph image. (top right) Radio dynamic spectrum from the Hiraiso radio spectrograph (HiRAS) showing no emission in the expected time window (marked by the red circle). (Bottom) The panels are similar to the top ones, except for an eruptive flare of the same size, but accompanied by a shock-driving CME and meter wave radio bursts of type II, type III, and type IV.

Figure 2 compares a confined flare with an eruptive flare of the same size (X1.5). The confined flare has a slightly shorter duration. The soft X-ray flares are due to energy deposition by electrons accelerated in the flare reconnection streaming towards the Sun. In both cases, 17 GHz microwave emission was observed (not shown), indicating sunward propagating nonthermal electrons. No radio emission is observed in the metric radio band indicating that there is no upward flow of nonthermal electrons in the confined flare case. On the contrary, the eruptive flare is associated with a shock-driving CME. The shock also manifests as an EUV wave as can be seen in the SDO/AIA 193 Å difference image. In addition, type II, type III, and type IV bursts can be found in the metric radio dynamic spectrum. While type III and type IV bursts are thought to be due to electrons accelerated in the flare reconnection region, the type II burst is due to electrons accelerated in the shock front. Furthermore, the eruptive event was associated with a large (215 pfu) SEP event (see Fig. 3). The SEP event, metric type II burst, and the IP type II burst are all due to the CME-driven shock. The triple plots in Fig. 3 available online are useful in identifying the underlying CME and flare for SEP events.

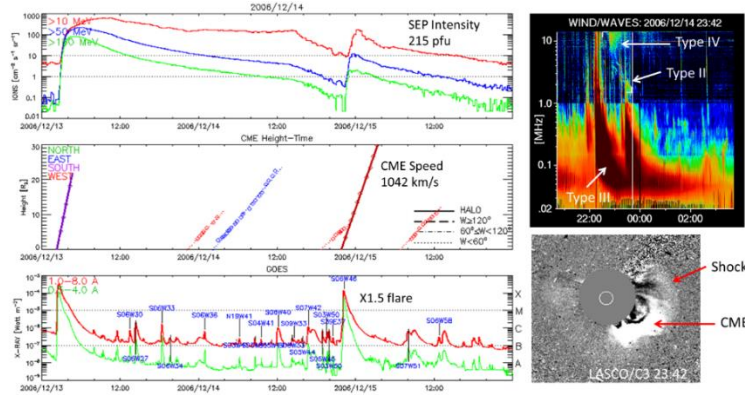

**Fig. 3**. (left) A composite plot showing the GOES SEP flux in pfu in three integral channels, the CME height-time plots for all CMEs occurring in a 3-day window, and the GOES soft X-ray light curve in two channels 1-8 Å and 0.5 – 4 Å, where the heliographic locations of the individual flares are indicated in blue. The X1.5 flare from S06W48 (cf. Fig. 2) is towards the end of 2006 December 14, associated with a halo CME (sky plane speed = 1042 km/s) and a large SEP event (intensity = 215 pfu). Plot from https://cdaw.gsfc.nasa.gov. (top right) A Wind/WAVES radio dynamic spectrum showing IP type II, Type II, and type IV bursts associated with the SEP event. (bottom right) SOHO LASCO CME and the shock towards the end of the type II burst.

CMEs originating from quiescent filament regions accelerate slowly and attain super-Alfvenic speeds at relatively large distances from the Sun, and the SEPs, if present, are purely from shock acceleration [38-39]. Also, when a CME-driven shock remains strong until 1 au and intercepted by a spacecraft with particle detectors, an energetic storm particle is observed. These particles are locally accelerated and hence represent purely shock-accelerated particles with little flare contribution. One of the outstanding problems in SEP studies is the relative contribution of particles originating from flare and shock sources in a given event [40-41]. Nevertheless, CME-driven shocks and type II bursts seem to be the best indicators of the large SEP events [42].

## 4  Intensity of SEP Events

Among the observed properties of SEP events, the intensity, energy spectra, and temporal evolution are the ones that decide the space weather impact of these events. These properties are controlled by the CME/shock characteristics as well as the conditions in the ambient medium in which the shock propagates.

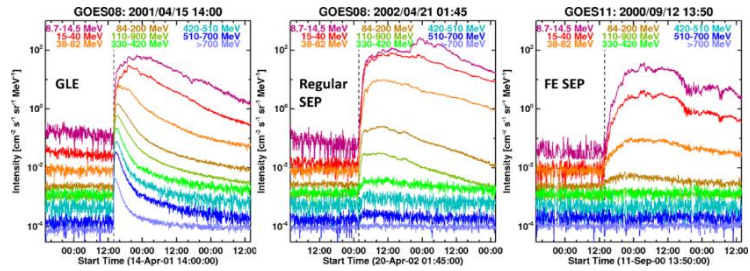

**Fig. 4**. Intensity vs. time profiles of three SEP events: (left) an SEP event with ground level enhancement (GLE) showing significant intensity increase in all energy channels, (middle) a regular SEP event with intensity signals in all channels except the highest one, (right) An SEP event due to a CME from a quiescent filament region (FESEP) that has intensity signal only in the lowest few channels. Clearly, spectral hardness decreases as one goes from left to right.

Figure 4 shows the time evolution of three large SEP events observed by GOES satellites. The 2001 April 15 event is a GLE event, indicating that particles are accelerated to GeV energies. The GOES light curves show a large peak in the >700 MeV channel. The 2002 April 21 event is a regular SEP event with significant particle intensities up to ~200 MeV. At higher energies, the intensity is relatively low, with no increase in the >700 MeV channel. The 2000 September 12 event has no intensity increase above 200 MeV. While the first two events are from an active region, the third event is from a quiescent filament region (FESEP event). One can infer that the spectra become progressively softer as one goes from GLE to regular SEP to FESEP events. The intensity and spectra of large SEP events are related to the properties of underlying CMEs as we illustrate in the next section.

The requirement of high CME speed for producing SEP events was reported a while ago [43-44]. A moderate correlation was found between CME speed and SEP intensity, which increased when shock speed was used in place of the CME speed [45]. It was pointed out that the SEP intensity varied by 3-4 orders of magnitude for a given CME speed [44], probably due to (i) spectral variation of individual events, and (ii) presence of seed particles in the heliosphere. These correlations are obtained using sky-plane speeds, which need to be revised using 3-D speeds. An issue related to the seed particles is the result of higher SEP intensity when CMEs happen in quick succession from the same active region [46]. We discuss some of these points in more detail below.

## 5      Which CMEs produce large SEP Events?

CMEs occur at the rate of one every other day during solar minima and several per day during solar Maxima. There have been about 30,000 CMEs recorded in the SOHO/LASCO CME catalog since 1996 when SOHO started routinely observing CMEs. This corresponds to an average rate of over 1000 CMEs per year. However, the number of CMEs that produce SEP events is much smaller – only a fraction of a percent. Figure 5 shows the speed, width, and solar source locations of CMEs that produced large SEP events since 1996. The speed and width are measured in the sky plane and the source locations are heliographic coordinates of the eruption. We see that the CME speed has an approximate normal distribution, with an average speed of ~1545 km/s. This is almost a factor of 4 higher than the average speed of the general population (~400 km/s). The lowest bin of the speed distribution (600 km/s) has ~10 CMEs, indicating that SEP-producing CMEs need to have above-average speed to drive a shock. The number of CMEs with speeds >1000 km/s drop rapidly and only a couple of CMEs have speeds exceeding 3000 km/s, limited by the amount of energy that can be stored/released in an active region. The width distribution shows that most (~82%) of the CMEs are halos. Halo CMEs are more energetic than regular CMEs and are very wide. The small number of non-halo CMEs have an average width of ~185° in the sky plane, which is quite large compared to the average width of regular CMEs, ~40°. Thus, only fast and wide CMEs possess enough energy to accelerate particles to energies >10 MeV. Figure 5c shows that the SEP-producing CMEs are concentrated in the latitude range ±30° corresponding to the active region belt. Furthermore, the source locations are heavily concentrated in the western hemisphere. This is because an Earth observer is magnetically connected to the western hemisphere (Parker spiral field lines) and energetic particles stream along those field lines to reach Earth. Energetic eastern hemispheric CMEs do produce SEP events but at locations behind the east limb. For example, the STEREO mission has detected large SEP events due to eastern hemispheric CMEs in Earth view, even though they did not produce a significant event at Earth [47].

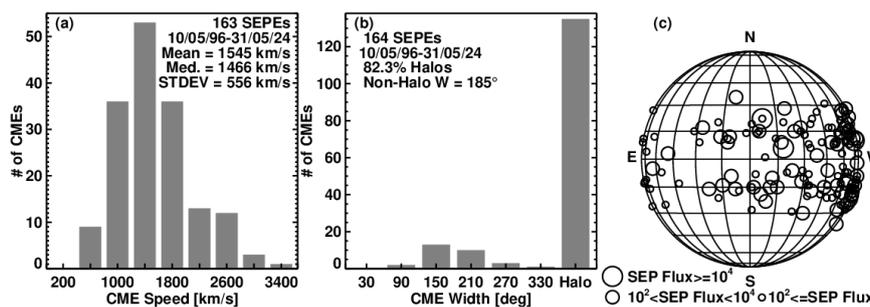

**Fig. 5**. Speed (a), width (b), and source location (c) distributions of CMEs associated with large SEP events observed by GOES since 1996, when SOHO started routinely observing CMEs. In (c), the size of the circle represents the SEP event size (peak intensity).

The speed used in Fig. 5 is an average within the coronagraph field of view (FOV, 2.5 – 32 Rs) measured in the sky plane. CMEs initially accelerate, reach a peak speed, and then decelerate [48]. The initial acceleration can range from a few m s$^{-2}$ to ~10 km s$^{-2}$. Some CMEs finish accelerating very close to the Sun, while others continue to accelerate over large distances. Therefore, one expects that the high mean speed in Fig. 5 can be attained at various heliocentric distances. The ambient magnetic field significantly declines with distance from the Sun and the ambient Alfven speed also declines following an initial peak around 3 Rs. While the shock strength depends on its speed and the ambient Alfven speed, the particle acceleration efficiency of shocks depends on the ambient magnetic field, and hence declines with heliocentric distance.

**Table 1**. Kinematics and spectral properties of events in Fig. 4

| SEP date | Ip pfu | $V_{CME}$ km/s | $a_{CME}$ m/s$^2$ | $V_{CMEi}$ km/s | $a_{flr}$ km/s$^2$ | Flare Size | Δt min | γ | Source Location |
|---|---|---|---|---|---|---|---|---|---|
| 2001/04/15 | 951 | 1199 | −35.9 | 1697 | 2.50 | X14.0 | 8 | 2.33 | S20W85 |
| 2002/04/21 | 2520 | 2393 | −1.39 | 2088 | 0.59 | X1.5 | 68 | 2.77 | S14W84 |
| 2000/09/12 | 321 | 1550 | +58.2 | 938 | 0.38 | M1.0 | 50 | 3.81 | S19W06 |

Table 1 lists CME properties of the three SEP events shown in Fig. 4. The average CME speed in Fig 5a is 1545 km/s, which is above, similar, and below the CME speed ($V_{CME}$) in the GLE event (2001 April 15: 1199 km/s), FESEP event (2000 September 12: 1550 km/s), and regular SEP event (2002 April 21: 2393 km/s), respectively. Clearly the >10 MeV SEP intensity (Ip, column 2) is not ordered by the sky-plane speed averaged over the coronagraph FOV. The average CME acceleration within the coronagraph FOV ($a_{CME}$, column 4) indicates different kinematics of the underlying CMEs (rapid deceleration in the GLE event, small deceleration in the regular SEP event, and continued acceleration in the FESEP event). The initial speed of CMEs ($V_{CMEi}$, column 5) computed from the first two height-time data points clearly orders the SEP intensities. A high initial speed implies shock formation close to the Sun. $V_{CMEi} > V_{CME}$ in the GLE event, whereas $V_{CMEi} < V_{CME}$ in the regular SEP and FESEP events. This means that the GLE CME finished accelerating close to the Sun, whereas the other two events continued to accelerate in the coronagraph FOV. This is indicated by the initial acceleration ($a_{flr}$, column 6) obtained as $V_{CME}/\Delta t$, where Δt (column 8) is the rise time of the associated soft X-ray (SXR) flare (column 7). The power law index of the fluence spectrum (γ, column 9) indicates the spectral hardness of the SEP events: the GLE event has the hardest spectrum and the FESEP event has the softest spectrum with the regular SEP event in-between. We see an inverse relationship between $a_{flr}$ and γ. Finally, the SXR flare size orders the SEP spectral hardness, but not the SEP intensity.

## 6 Shock Formation heights, SEP spectra, and CME kinematics

It is well known that the earliest indicator of a CME-driven shock is the onset time of a type II radio burst. Type II bursts typically start at a frequency of ~150 MHz. For a fundamental plasma emission, this frequency corresponds to an electron density of ~2.8×10$^8$ cm$^{-3}$, which typically occurs at a heliocentric distance of ~1.5 Rs [49]. There are bursts which start at higher frequencies (i.e., closer to the Sun) [50-53] or lower frequencies (i.e., farther from the Sun) [54] depending on the local plasma conditions and the CME kinematics. CMEs accelerating impulsively and attaining high speeds close to the Sun (within a couple of Rs) result in high-energy SEPs. This is because the particle acceleration efficiency of shocks is higher when the ambient magnetic field is higher [55]. Once a shock forms, it typically takes several minutes before the high-energy particles are released from the shock. For instance, in GLE events, the shock formation height is typically ~1.5 Rs and the particle release occurs when the CME is at a height of ~2 – 3 Rs [52,56]. Thus, the CME initial acceleration, initial speed, shock formation height, and height of particle release are the key parameters that are of interest to space weather research connected to energetic particles (see Fig. 6).

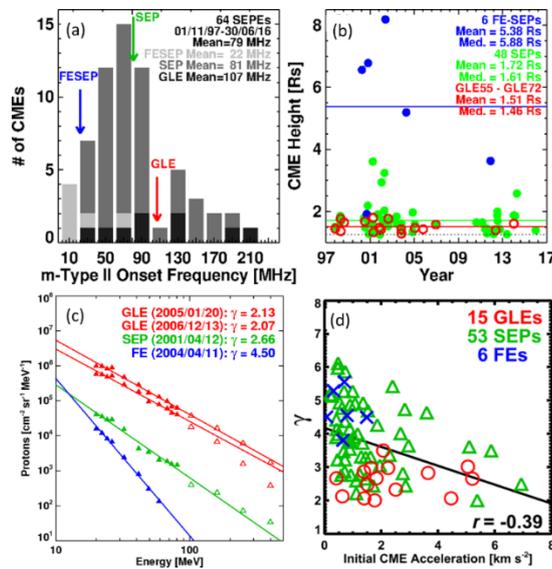

**Fig. 6.** (a) Histograms of starting frequencies of type II radio bursts associated with GLE events (dark bars), regular SEP events (dark gray bars), and FESEP events (light gray bars). The means of the distributions are shown on the plot and pointed by arrows. (b) Shock formation heights obtained as the CME leading-edge height at the time of the type II burst onset for the three groups of SEP events. (c) Representative 10-100 MeV fluence spectra of two GLE events (red), a regular SEP event (green), and an FESEP event (blue). The spectral indices (γ) are noted on the plot. (d) Variation of the fluence spectral index as a function of CME initial acceleration distinguished by the event group. The initial acceleration was computed as $a_{flr} = V_{CME}/\Delta t$, from the average the average CME speed ($V_{CME}$) in the coronagraph field of view and the flare rise time $\Delta t$. The correlation is moderate but statistically significant (the Pearson critical correlation coefficient is ~0.376 for p = 0.0005, the probability that the reported correlation is by chance).

Figure 6a shows that Type II bursts occur at highest frequencies in GLE events (~107 MHz), followed by regular SEP events (~81 MHz) and FESEP events (22 MHz). A similar pattern can be recognized in the shock formation height obtained as the CME leading-edge height at the time of type II burst onset (see Fig. 6b). Although there is some overlap, the mean shock formation height is ~1.51 Rs for GLE events, 1.72 Rs for regular SEP events, and 5.38 Rs for FESEP events. Accordingly, the hardness of the 10-100 MeV fluence spectrum progressively decreases from GLE to regular SEP to FESEP events as illustrated in Fig. 6c. Power-law fits to the particle fluence (Fp) data of the form Fp ~E$^{-\gamma}$ show that the GLE events have the hardest spectra ($\gamma$ ~2.0), regular SEP events have moderate hardness ($\gamma$ ~2.66), and the FESEP events have the softest spectra ($\gamma$ ~4.50). The progressive spectral hardness is further illustrated in Fig. 6d: the spectral index is negatively correlated with the CME initial acceleration with a correlation coefficient of ~ –0.39.

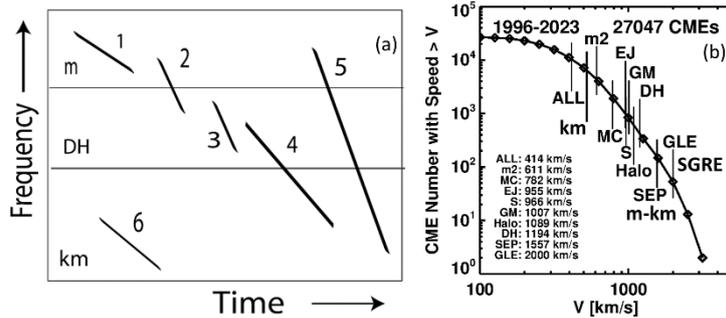

**Fig. 7**. (a) Type II radio bursts in a schematic dynamic spectrum: pure metric (m, 1), metric to decameter-hectometric (m-DH, 2), pure DH (3), DH-kilometric (DH-km, 4), m-km (5), and pure km (6). Cases 2-5 are sometimes combined as DH for simplicity. (b) average speeds of various CME populations associated with metric (m2), km, DH, and m-km type II bursts; other populations included for comparison are those associated with magnetic clouds (MC), non-cloud ejecta (EJ), interplanetary shocks (S), intense geomagnetic storms (GM), SEP events (SEP), ground level enhancements (GLE) in SEP events, and sustained gamma-ray emission (SGRE) events; the average speed of halo CMEs is also shown for comparison (Halo).

## 7 Shock Formation heights, SEP spectra, and CME kinematics

While the starting frequency of type II bursts indicate the height of shock formation, the ending frequency indicates the shock strength. Type II bursts have been observed to start and end at various wavelength domains from meter waves (m) to kilometer waves (km) with the intervening decameter-hectometric (DH) waves (see Figure 7). Type II bursts can start and end in the m domain and are associated with CMEs with an average speed of ~600 km/s (case 1 in Fig. 7a). If we combine cases 2, 3, and 4, the underlying CMEs have an average speed of ~1200 km/s. CMEs producing type II emission components at all wavelengths from m to km (case 5) have the highest average speed (~1500 km/s). Finally, CMEs producing type II bursts only in the km domain (case 6) have the lowest average speed (~500 km/s). These average values are marked and compared with the speeds of other energetic CME populations in Fig. 7b. CMEs in all cases except case 6 have an average deceleration within the coronagraphic field of view, while those in case 6 continue to accelerate and become super-Alfvenic only at tens of Rs from the Sun and hence they produce type II bursts only in the km domain. The hierarchical relationship between CME speed (or kinetic energy) and the wavelength range of type II bursts is an indication of the close connection between shock strength and duration of particle acceleration [54,57]. The m-km type II bursts indicate the strongest shocks that propagate to 1 au and even beyond.

The fact that type II bursts extending beyond the metric domain indicate stronger shocks has been illustrated in terms of SEP association [58]: only 25% of the western hemispheric m type II bursts (similar to case 1 in Fig, 7a) were associated with large SEP events, increasing to 90% when the metric type II bursts were accompanied by DH type II bursts. Their DH type II bursts include all type II events with ending frequency below 14 MHz. As noted, m-km type II bursts are associated with very fast CMEs, so they must have the best association with SEP events. Figure 8 compares the speed, width, and residual acceleration of CMEs associated with SEP events and m-km type II bursts. The CME properties are the same because the same shock accelerates electrons to produce type II bursts and ions that constitute SEP events.

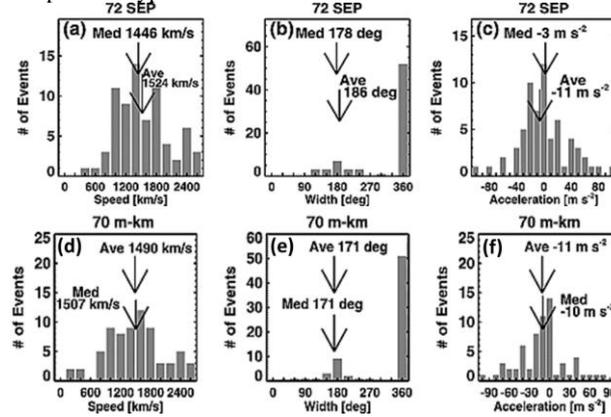

**Fig. 8.** Distributions of CME speed, width, and residual acceleration of CMEs associated with large SEP events (top) and type II radio bursts with emission components from metric to kilometric (m-km) wavelengths (bottom). Note that the CME properties are nearly the same.

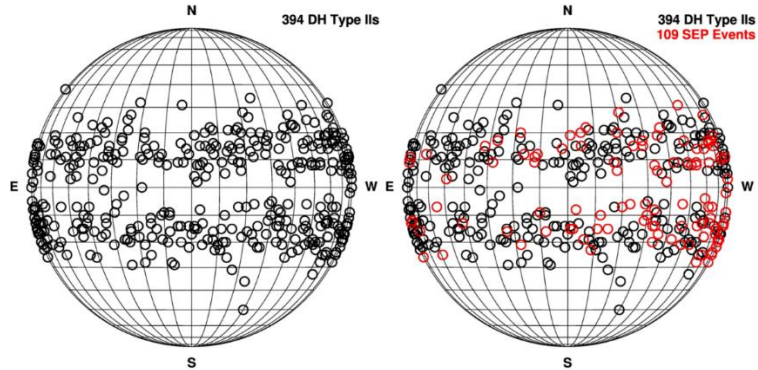

**Fig. 9.** Solar source locations of CMEs that produced a type II radio burst observed by the Wind/WAVES instrument (left). All type II bursts with an emission component in the DH domain are included (cases 2-5 in Fig. 7a). Heliographic coordinates of 394 CME sources are plotted. The grid spacing is 10 degrees in longitude and latitude. The sources of 109 CMEs are plotted in red indicating that they are associated with large SEP events (right). From [60].

The SEP association rate of DH type II bursts (cases 2-5 in Fig. 7a combined) from the western hemisphere steadily increases with CME speed, attaining 100% for $V_{CME} \geq$ 1800 km/s [59], consistent with the results in Figs. 7-8. The SEP-DH type II burst association is further illustrated in Fig. 9, which plots the solar sources of CMEs associated with DH type II bursts. We see that the source locations of type II bursts are clustered around two latitude ranges corresponding to the active region belt, and uniformly distributed in longitude (type IIs are electromagnetic emission and hence are not affected by the IP magnetic field). Type II bursts originating mainly from the western longitudes are associated with SEP events because of the required magnetic connectivity to detect an SEP event at Earth. A small fraction (~21%) of DH type II bursts from the eastern hemisphere do have SEP association. A closer look at these events finds that they are associated with extremely fast and CMEs [60]. Such energetic CMEs drive extended shocks whose flanks that have magnetic connectivity to Earth.

What are the space weather implications of the type II burst – SEP association? Type II bursts can isolate the small number of energetic CMEs that drive a shock. A western hemispheric type II burst at the Sun indicates a high probability of occurrence of a large SEP event. Ten MeV protons have a speed of ~43470 km/s, so they can travel to Earth in ~75 min (assuming a Parker spiral path length of ~1.3 AU) if they are not scattered significantly during propagation. Normalizing to the arrival of electromagnetic emission at Earth, this represents an advance warning of about an hour. However, the higher energy particles travel much faster and hence the type II bursts cannot provide meaningful warning. For example, GeV particles (in GLE events) arrive at Earth only about ~3 minutes behind the electromagnetic emissions. The onset times of a type II burst or an SEP event may still be useful because spacecraft anomalies peak a couple of days after the onset of SEP events [19].

The scatter-free propagation noted above is not realistic because the real solar wind has turbulence superposed on the Parker spiral field lines. Therefore, a proper understanding of the transport of energetic particles in the ambient solar wind is needed to predict their properties at Earth. Solar wind properties are typically measured at 1 au, so a model is needed to characterize the solar wind/ambient medium from 1 au to the shock vicinity where the particles are released. This model needs to be coupled to another model that describes the transport of the particles. There are several such models developed recently with this general procedure of combining a solar wind model with a transport model [61-65]. A full list of SEP models, their current state of development, and validations can be found in [66].

## 8  Solar Cycle Variation of SEP Events

Figure 10 illustrates the solar cycle (SC) variation of peak SEP intensity (from GOES) and the source latitudes during five solar cycles since 1976. A higher concentration of SEP events near solar maxima is evident, although a significant number of events can be found during the declining phases (more pronounced in cycles 21 and 23). All cycles show events with NOAA scales at least S3 (i.e., $10^3$ pfu, see section 2), except for the partial cycle 25. SCs 22 and 23 have many S4 ($10^4$ pfu) events, but SC21 and 23 have S3 events, but no S4 events. Even though SC 21 has the highest sunspot number (SSN), the SEP intensities are relatively low, indicating that SSN alone is not an indicator of SEP intensities. This is because some large active regions are CME poor and hence are not associated with SEP events. Furthermore, some active regions emerge close to the end of a cycle, resulting in large SEP events. For example, there were four large SEP events during February-May of 1986 from ARs 4711, 4713, 4717, and 4717. AR 4711 was reported to be very active during its disk passage (e.g., [67] and references therein).

Figure 10b shows that SEPs occur in clusters in both northern and southern hemispheres and on either side of the SSN maximum. The GLE events have a distribution similar to that of SEP events. In cycles 21-23, there were about a dozen GLE events in each cycle. The GLE occurrence rate drastically dropped in SC 24 that had only two GLE events. The lack of high-energy SEP events in SC 24 has been attributed to the weakened heliospheric state in that cycle [68]. SC 25 is in its maximum phase but has only two GLEs in the first 4.5 years indicating that this cycle is similar to SC 24.

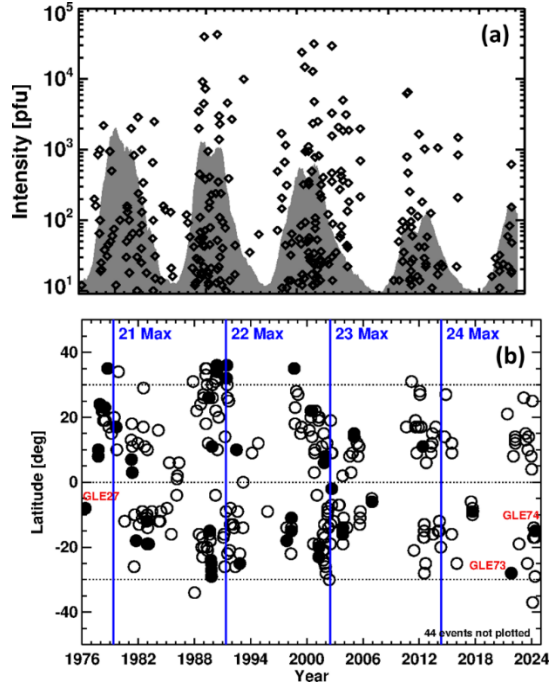

**Fig. 10**. (a) Maximum intensity (in pfu) of large SEP events as a function of time as observed by GOES in the >10 MeV integral channel. The Y-axis corresponds to the NOAA radiation storm scale S1-S5 (see section 2). The sunspot number is shown in gray for reference (from WDC-SILSO, Royal Observatory of Belgium, Brussels). The maximum intensity often occurs when the associated shock arrives at 1 au indicating that the peak is due to energetic storm particle (ESP) events. (b) Latitude vs. time plot of the solar source locations of large SEP events that occurred since 1976. Filled circles denote GLE events. Three GLE events at the beginning and end are identified. The blue lines mark the time of maximum of cycles 21-24.

## 9  Summary

SEPs result from fundamental physical processes that convert free energy stored in solar magnetic fields to the kinetic energy of particles in the corona and IP medium. SEPs are also of great practical importance since they pose radiation hazard to humans and their technology in space. They degrade solar panels of satellites, cause satellite anomalies (deep dielectric discharges, single event upsets) in high and low latitude satellites, result in polar cap absorption of radio waves, modify Van Allen belt particle population, deplete ozone in Earth's atmosphere, and provide radiation dose to crew and passengers in aircraft flying in polar routes. Radio bursts are electromagnetic phenomena accompanying solar eruptions that result in SEPs. Long-duration type III bursts are indicative of electron acceleration in solar flares associated with
CMEs. Low-frequency type IV bursts are indicative of flare-accelerated electrons in post eruption arcades and CME flux ropes. Type II bursts are due to electrons accelerated in CME-driven shocks that also energize ions and hence are a key electromagnetic signature of SEP events. Fast and wide CMEs that attain high speeds close to the Sun are responsible for the large SEP events. The intensity, spectra, and time evolution SEP events determine the ensuing space weather impact, but it is difficult predict these because of the high energies involved. These SEP properties are closely related to the properties of underlying CMEs such as early kinematics, shock formation distance from the Sun, and magnetic connectivity of the source region to the observer.

**Acknowledgments.** Work supported by NASA's Living With a Star Program and the STEREO project.